\documentclass{article}
\usepackage[psamsfonts]{amssymb}
\usepackage{amsmath}
\usepackage{cite}
\usepackage{bm}% bold math
\usepackage{graphicx}
\usepackage{yhmath}
\usepackage{bbm}

\def\d{\mathrm{d}}

\def\id{\mathbf{1}}

\def\ir{\mathrm{i}}

\def\tr{\mathrm{tr}}
\def\Z{\mathcal{Z}}
\def\I{\mathrm{I}}
\begin{document}
\begin{titlepage}
\noindent{\large\textbf{Statistical mechanics of free particles on
space with Lie type noncommutativity}}

\vspace{\baselineskip}

\begin{center}
{
Ahmad~Shariati~{\footnote {shariati@mailaps.org}}}\\
\vskip 0.1cm
Mohammad~Khorrami~{\footnote {mamwad@mailaps.org}}\\
\vskip 0.1cm
Amir~H.~Fatollahi~{\footnote {ahfatol@gmail.com}}\\
\vskip 0.5cm \textit{ Department of Physics, Alzahra University,
Tehran 1993891167, Iran }
\end{center}
\vspace{\baselineskip}
\begin{abstract}
\noindent Effects of Lie type noncommutativity on thermodynamic properties of a
system of free identical particles are investigated. A definition for finite volume of
the configuration space is given, and the grandcanonical partition function in the
thermodynamic limit is calculated. Two possible definitions for the pressure are
discussed, which are equivalent when the noncommutativity vanishes. The thermodynamic observables
are extracted from the partition function. Different limits are discussed where either
the noncommutativity or the quantum effects are important. Finally specific cases are discussed
where the group is SU(2) or SO(3), and the partition function of a nondegenerate gas is calculated.
\end{abstract}
\end{titlepage}

\section{Introduction}
Noncommutative spacetime is recognized as the space whose coordinate
operators do not commute. In the simplest case of canonical noncommutativity
(the so-called Groenewold-Moyal space) the coordinates satisfy
\begin{equation}\label{kfs.1}
[\hat{x}_\mu,\hat{x}_\nu]=\ir\,\theta_{\mu\,\nu}\,\id,
\end{equation}
in which $\theta$ is an antisymmetric constant tensor and $\id$
as the unit operator. The theoretical and phenomenological
implications of such noncommutative coordinates have been extensively
studied during last decade \cite{reviewnc}, once it was understood
that the longitudinal directions of D-branes in the presence of a
constant B-field background appear to be noncommutative, as seen by
the ends of open strings \cite{9908142,99-2,99-3,99-4}.

One direction to extend studies on noncommutative spaces is to
consider spaces where the commutators of the coordinates are not
constants. Examples of this kind are the noncommutative cylinder
and the $q$-deformed plane (the Manin plane \cite{manin}, the so-called
$\kappa$-Poincar\'{e} algebra \cite{luk} (see also \cite{majid,ruegg,amelino,kappa}),
and linear noncommutativity of the Lie algebra type
\cite{synder} (see also \cite{wess,sasak}). In the latter the dimensionless spatial
positions operators satisfy the commutation relations of a Lie
algebra:
\begin{equation}\label{kfs.2}
[\hat{x}_a,\hat{x}_b]= f^c{}_{a\, b}\,\hat{x}_c,
\end{equation}
where $f^c{}_{a\,b}$'s are structure constants of a Lie algebra.
One example of this kind is the algebra SO(3), or SU(2). A special
case of this is the so called fuzzy sphere \cite{madore} (see also \cite{presnaj}),
where an irreducible representation of the position operators is
used which makes the Casimir of the algebra,
$(\hat{x}_1)^2+(\hat{x}_2)^2+(\hat{x}_3)^2$, a multiple of the
identity operator (a constant, hence the name sphere). One can
consider the square root of this Casimir as the radius of the
fuzzy sphere. This is, however, a noncommutative version of a
two-dimensional space (sphere).

In \cite{0612013,fakE1,fakE2} a model was introduced in which the
representation was not restricted to an irreducible one, instead
the whole group was employed. In particular the regular
representation of the group was considered, which contains all
representations. As a consequence in such models one is dealing
with the whole space, rather than a sub-space, like the case of
fuzzy sphere as a 2-dimensional surface. In \cite{0612013} basic
ingredients for calculus on a linear fuzzy space, as well as basic
notions for a field theory on such a space, were introduced. In
\cite{fakE1, fakE2} basic elements for calculating the matrix elements
corresponding to transition between initial and final states,
together with the explicit expressions for tree and one-loop
amplitudes were given. It is observed that models based on
Lie algebra type noncommutativity enjoy three features:
\begin{itemize}
\item They are free from any ultraviolet divergences if the group
is compact.
\item There is no momentum conservation in such
theories.
\item In the transition amplitudes only the so-called
planar graphs contribute.
\end{itemize} The reason for latter is that the non-planar graphs
are proportional to $\delta$-distributions whose dimensions are
less than their analogues coming from the planar sector, and so
their contributions vanish in the infinite-volume limit usually
taken in transition amplitudes \cite{fakE2}. The consequence of
different mass-shell condition of these kinds theory was explored
in \cite{skf}.

In \cite{kfs} the classical mechanics defined on a space with
SU(2) fuzziness was studied. In particular, the Poisson structure
induced by noncommutativity of SU(2) type was investigated, for
either Cartesian or Euler parameterization of SU(2) group. The
consequences of SU(2)-symmetry in such spaces on integrability,
was also studied in \cite{kfs}. In \cite{fsk} the quantum mechanics
on a space with SU(2) fuzziness was examined. In particular,
the commutation relations of the position and momentum
operators corresponding to spaces with Lie-algebra
noncommutativity in the configuration space, as well as
the eigen-value problem for the SU(2)-invariant systems were studied.

The consequences of the noncommutativity of space on thermodynamical
properties have been explored in a statistical mechanics and filed
theoretical approach. In \cite{scho} the thermodynamics of
a fermion gas is considered in a space with noncommutativity of
canonical type. In \cite{shin} the thermal effects were considered
on fuzzy sphere or on spaces which are the result of the direct product of
a Minkowski with a fuzzy sphere. The potential importance of such
kind of studies can be understood once one mentions that it is quite expected
that the noncommutative effects would be detectable only in such high energy
processes which could be possible only in very hot seconds of the early
universe.

The purpose of this work is to explore the effect of Lite type noncommutativity
on thermodynamics of physical systems. In particular we consider the
free particles which obey boson, fermion and classical statistics.
First, we give a recipe to give a practical meaning to ``a finite volume"
in a space with Lie type fuzziness. Second, in thermodynamical limit,
we give the proper expression for the grand canonical partition function
of free gas. It is explained how one can define in two inequivalent ways
the pressure. The different limits were considered in calculation of
the thermodynamical quantities.

\section{The Hilbert space}
The Hilbert space is defined as the space of $L^2$-distributions
defined on the group manifold, where the integration measure is 
the Haar measure of the group. The group is assumed to be unimodular, 
so that the left- and right-Harr measures coincide, and also equal to 
its identity component, so that the exponential map is surjective. A
completeness relation for the orthonormal kets $|U\rangle$ can be
written like
\begin{equation}\label{04.101}
\int\d U\;|U\rangle\langle U|=\id.
\end{equation}
The element $|U\rangle$ corresponds the distribution $\delta_U$
with
\begin{equation}\label{04.1001}
\delta_U(U')=\delta(U^{-1}\,U'),
\end{equation}
where $\delta$ is the Dirac distribution:
\begin{equation}\label{04.1002}
\int\d U'\;\delta(U^{-1}\,U')\,f(U')=f(U).
\end{equation}
It is clear that $|U\rangle$ does not belong to the Hilbert space,
but to an extension of it. Yet the elements of the Hilbert space 
can be expanded in terms of $|U\rangle$'s.

The group element $U$ itself is written as a function of the
coordinates $\hat{k}^a$ according to
\begin{equation}\label{04.102}
U(\hat{\mathbf{k}}):=[\exp(\hat{k}^a\,\hat{x}_a)]\,U(\mathbf{0}),
\end{equation}
where $U(\hat{\mathbf{k}})$ is the group element corresponding to
the coordinates $\hat{\mathbf{k}}$, $U(\mathbf{0})$ is the
identity, and $\exp(\hat{x})$ is the flux corresponding to the
vector field $\hat{x}$. The set of $\hat{x}_a$'s is a basis for
the left-invariant vector fields. The action of $L_{\hat{x}_a}$
(the Lie derivative corresponding to the vector field $\hat{x}_a$)
on an arbitrary scalar function $F$ can be written like
\begin{equation}\label{04.103}
L_{\hat{x}_a}(F)=\hat{x}_a{}^b\,\frac{\partial\,F}{\partial\hat{k}^b},
\end{equation}
where $\hat{x}_a{}^b$'s are scalar functions, and satisfy
\begin{equation}\label{04.104}
\hat{x}_a{}^b(\hat{\mathbf{k}}=\mathbf{0})=\delta_a^b.
\end{equation}
The vector fields $\hat{x}_a^{\mathrm{R}}$ are defined as
right-invariant vector fields coinciding with their left-invariant
analogues at the identity of the group:
\begin{equation}\label{04.105}
\hat{x}_a^{\mathrm{R}}(\hat{\mathbf{k}}=\mathbf{0})=
\hat{x}_a(\hat{\mathbf{k}}=\mathbf{0}).
\end{equation}
Finally, the generators of the adjoint action are defined as
\begin{equation}\label{04.106}
\hat{J}_a:=\hat{x}_a-\hat{x}_a^{\mathrm{R}}.
\end{equation}
Dimensionalizing these as
\begin{align}\label{04.107}
p^a&:=(\hbar/\ell)\,\hat{k}^a,\\
\label{04.108} x_a&:=\ir\,\ell\,\hat{x}_a,\\
\label{04.109} x_a{}^b(\mathbf{p})&:=
\hat{x}_a{}^b[(\ell/\hbar)\,\mathbf{p}],\\
\label{04.110} J_a&:=\ir\,\hbar\,\hat{J}_a,
\end{align}
where $\ell$ is a constant of dimension length, one arrives at the
following commutation relations \cite{kfs,fsk}.
\begin{align}\label{04.111}
[p^a,p^b]&=0,\\
\label{04.112} [x_a,p^b]&=\ir\,\hbar\,x_a{}^b,\\
\label{04.113} [x_a,x_b]&=\ir\,\ell\,f^c{}_{a\,b}\,x_c,\\
\label{04.114} [J_a,x_b]&=\ir\,\hbar\,f^c{}_{a\,b}\,x_c,\\
\label{04.115} [p^c,J_a]&=\ir\,\hbar\,f^c{}_{a\,b}\,p^b,\\
\label{04.116} [J_a,J_b]&=\ir\,\hbar\,f^c{}_{a\,b}\,J_c,
\end{align}
where $x_a$'s and $p^b$'s are the coordinate and momentum
operators, respectively.
\section{The partition function}
The Hamiltonian corresponding to free particles in a space of 
infinite volume is a function of only momenta. Denoting this by $H$, one can
define another Hamiltonian corresponding to free particles in a
space of finite volume. In the case of Lie-algebra type noncommutative
spaces, finiteness of the volume of the space can be implemented by restricting
the representations of the coordinate operators. So, corresponding
to any operator $Q$ acting on the Hilbert space corresponding to
the space of infinite volume, one constructs the operator $Q_V$ acting on
the Hilbert space corresponding to a space of of volume $V$ through
\begin{equation}\label{04.117}
Q_V:=\Pi_V\,Q\,\Pi_V,
\end{equation}
where $\Pi_V$ is the Hermitian projection operator the image of
which is the subspace of the Hilbert space corresponding to the
desired representations. The aim is to find the partition function
in the thermodynamic limit that the volume of the space becomes 
infinite (so that $\Pi_V$ tends to $\id$).

Denoting the grand canonical partition function of a system of
free particles in a volume $V$ by $\Z(V)$, one has
\begin{equation}\label{04.118}
\ln\Z(V)=-\frac{1}{s}\,\tr\{\ln[1-s\,z\,\exp(-\beta\,H_V)]\},
\end{equation}
where $H$ is the one particle Hamiltonian, $z$ is the fugacity,
related to the temperature and the chemical potential $\mu$ through
\begin{equation}\label{04.119}
z:=\exp\left(\frac{\mu}{k_\mathrm{B}\,T}\right),
\end{equation}
and
\begin{equation}\label{04.120}
\beta:=\frac{1}{k_\mathrm{B}\,T}.
\end{equation}
$k_\mathrm{B}$ is the Boltzmann's constant, and $T$ is the
absolute temperature. Bosons, fermions, and the fictitious classical particles
(classons) correspond to the $s=+1$, $s=-1$, and the limit $s\to 0$, respectively .
Also note that any identity operator in the right hand side of (\ref{04.118}) is the identity
operator of the restricted Hilbert space.

To calculate the right hand side of (\ref{04.118}) in the
thermodynamic limit, one notes that
\begin{align}\label{04.121}
\tr (H_V)^j=&\int\d U\;\langle
U|\Pi_V\,(\Pi_V\,H\,\Pi_V)^j\,\Pi_V|U\rangle,\nonumber\\
=&\int\d U\,\d U_1\cdots\d U_j\;\langle U|\Pi_V\,H|U_1\rangle
\langle U_1|\Pi_V\,H|U_2\rangle\cdots\nonumber\\
&\times\langle U_{j-1}|\Pi_V\,H|U_j\rangle\langle
U_j|\Pi_V|U\rangle,\nonumber\\
=&\int\d U\,\d U_1\cdots\d U_j\;\langle U|\Pi_V|U_1\rangle
\langle U_1|\Pi_V|U_2\rangle\cdots\nonumber\\
&\times\langle U_{j-1}|\Pi_V|U_j\rangle\langle
U_j|\Pi_V|U\rangle\,E(U_1)\cdots E(U_j),
\end{align}
where $E(U)$ is the eigenvalue of $H$ corresponding the
eigenvector $|U\rangle$. In the thermodynamic limit the projection
$\Pi_V$ tends to the identity, hence its matrix elements tend to
the delta distribution. So in the right hand side of
(\ref{04.121}), up to the leading order one can substitute
$E(U_i)$ by $E(U)$, arriving at
\begin{align}\label{04.122}
\tr (H_V)^j=&\int\d U\,\d U_1\cdots\d U_j\;\langle
U|\Pi_V|U_1\rangle
\langle U_1|\Pi_V|U_2\rangle\cdots\nonumber\\
&\times\langle U_{j-1}|\Pi_V|U_j\rangle\langle
U_j|\Pi_V|U\rangle\,[E(U)]^j,\nonumber\\
=&\int\d U\;\langle U|\Pi_V|U\rangle\,[E(U)]^j.
\end{align}
One has
\begin{equation}\label{04.123}
|U(\hat{\mathbf{k}})\rangle=\exp[\hat
k^a\,x_a/(\ir\,\ell)]\,|U(\mathbf{0}\rangle,
\end{equation}
(where $\hat k^a$'s are numbers not operators). Assuming that the
representations kept corresponding to the finite-volume space are
determined by only the value of their corresponding Casimirs, it
is seen that $\Pi_V$ is a function of only Casimirs. So it turns
out that $\Pi_V$ commutes with the coordinate operators, from
which one arrives at
\begin{equation}\label{04.124}
\langle U|\Pi_V|U\rangle= \langle
U(\mathbf{0})|\Pi_V|U(\mathbf{0})\rangle.
\end{equation}
So
\begin{align}\label{04.125}
\tr (H_V)^j=& \langle U(\mathbf{0})|\Pi_V|U(\mathbf{0})\rangle \,
\int\d U\;[E(U)]^j.
\end{align}
Using this, one arrives at
\begin{equation}\label{04.126}
\frac{1}{\langle U(\mathbf{0})|\Pi_V|U(\mathbf{0})\rangle}
\,\ln\Z(V)=-\frac{1}{s}\,\int\d
U\;\ln\{1-s\,z\,\exp[-\beta\,E(U)]\}.
\end{equation}
From (\ref{04.125}) it is seen that the product of $\langle
U(\mathbf{0})|\Pi_V|U(\mathbf{0})\rangle$ and the Haar measure is
independent of the normalization choice for the Haar measure, so
that (\ref{04.126}) is in fact independent of the normalization
choice for the Haar measure. From now on, the normalization of the
haar measure is chosen so that
\begin{equation}\label{04.127}
\lim_{\hat k\to 0}\left[\frac{\d^D\hat k}{(2\,\pi\,\ell)^D\,\d
U}\right]=1,
\end{equation}
which is equivalent to
\begin{equation}\label{04.128}
\lim_{\hat k\to 0}\left[\frac{\d^D p}{(2\,\pi\,\hbar)^D\,\d
U}\right]=1,
\end{equation}
where $D$ is the dimension of the group, and
\begin{equation}\label{04.129}
\hat k:=\sqrt{\delta_{a\,b}\,\hat{k}^a\,\hat{k}^b}.
\end{equation}
In the commutative limit ($\ell\to 0$), the limiting cases of
(\ref{04.127}) or (\ref{04.128}) always apply, and the denominator
in the left hand side of (\ref{04.126}) is the volume of the
system. So one defines the volume of the noncommutative system as
\begin{equation}\label{04.130}
V:=\langle U(\mathbf{0})|\Pi_V|U(\mathbf{0})\rangle.
\end{equation}
One can explicitly check the meaning of this definition for the
groups SU(2) and SO(3). Suppose the spin of largest representation
which is kept is $\mathcal{J}$. Keeping in mind that the
representation with spin $j$ (not greater than $\mathcal{J}$) has
dimension $(2\,j+1)$ and appears $(2\,j+1)$ times, it is seen that
\begin{equation}\label{04.131}
\tr\Pi_V=\sum_{j=0}^\mathcal{J}(2\,j+1)^2,
\end{equation}
which (up to leading order) results in
\begin{equation}\label{04.132}
\tr\Pi_V=\begin{cases}\frac{8}{3}\,{\mathcal{J}}^3,&\mathrm{SU(2)}\\
\frac{4}{3}\,{\mathcal{J}}^3,&\mathrm{SO(3)}\end{cases}.
\end{equation}
The difference between these two groups is that for SU(2) only
$(2\,j)$ should be an integer, while for SO(3) the value of $j$
itself should be an integer. For these groups, the Haar measure
reads
\begin{equation}\label{04.133}
\d U=\frac{4}{(2\,\pi\,\ell)^3}\,\sin^2\frac{\hat k}{2}\,\d\hat
k\,\d\Omega,
\end{equation}
where $\Omega$ is the angular part of the spherical coordinates of
$\hat{\mathbf{k}}$, and
\begin{equation}\label{04.134}
\begin{cases}0\leq\hat k\leq 2\,\pi,&\mathrm{SU(2)}\\
0\leq\hat k\leq \pi,&\mathrm{SO(3)}\end{cases},
\end{equation}
so that
\begin{equation}\label{04.135}
\int\d U=\frac{4}{(2\,\pi\,\ell)^3}\,\begin{cases}16\,\pi^2,&\mathrm{SU(2)}\\
8\,\pi^2,&\mathrm{SO(3)}\end{cases}.
\end{equation}
Using (\ref{04.125}) with $j=0$, one arrives at
\begin{equation}\label{04.136}
\langle U(\mathbf{0})|\Pi_V|U(\mathbf{0})\rangle=
\frac{4\,\pi}{3}\,(\mathcal{J}\,\ell)^3,
\end{equation}
which is the volume of a sphere of radius $(\mathcal{J}\,\ell)$.
One also notices that the largest eigenvalue of
$\mathbf{x}\cdot\mathbf{x}$ is equal to
$\mathcal{J}\,(\mathcal{J}+1)\,\ell^2$, which is (to the leading
order) the square of the same radius $(\mathcal{J}\,\ell)$.
\section{Thermodynamic quantities}
Starting from the grand canonical partition function
(\ref{04.126}), one can easily obtain the number density and the
internal energy density in a manner similar to the commutative
case:
\begin{align}\label{04.137}
\frac{\mathcal{N}}{V}&=z\,\frac{\partial}{\partial
z}\left(\frac{\ln\mathcal{Z}}{V}\right),\nonumber\\
&=\int\d U\;\frac{z\,\exp[-\beta\,E(U)]}
{1-s\,z\,\exp[-\beta\,E(U)]},
\end{align}
and
\begin{align}\label{04.138}
\frac{\mathcal{E}}{V}&=-\frac{\partial}{\partial\beta}
\left(\frac{\ln\mathcal{Z}}{V}\right),\nonumber\\
&=\int\d U\;\frac{z\,\exp[-\beta\,E(U)]}
{1-s\,z\,\exp[-\beta\,E(U)]}\,E(U),
\end{align}
where $\mathcal{N}$ and $\mathcal{E}$ are the expectation values
of the number of particles and the energy of the system,
respectively. One can also define a number density in the phase
space like
\begin{equation}\label{04.139}
n(U):=\frac{z\,\exp[-\beta\,E(U)]} {1-s\,z\,\exp[-\beta\,E(U)]},
\end{equation}
so that
\begin{align}\label{04.140}
\frac{\mathcal{N}}{V}&=\int\d
U\;n(U),\nonumber\\
\frac{\mathcal{E}}{V}&=\int\d U\;n(U)\,E(U).
\end{align}
These look exactly similar to the corresponding expressions in the
commutative case, apart from the difference in the integration
measure and the functions involved. Regarding the pressure,
however, there arises a new concept. The point is that the way to
change the volume of the system is not unique. One can change the
noncommutativity length $\ell$, or the largest representation
involved. In the commutative limit $(\ell\to 0)$, the energy
function can be written so that it depends on only $p$, or the
combination $(\hat k/\ell)$. The same is true for the integration
measure and the integration region (which is infinite) in the
right hand side of (\ref{04.126}). So the grand canonical
partition function dependence on the representation and $\ell$ is
only through the volume $V$, and as $V$ is proportional to
$\ell^D$, it is seen that
\begin{equation}\label{04.141}
\frac{\ln\mathcal{Z}}{V}=\frac{\ell^D}{V}\,\frac{\partial
\ln\mathcal{Z}}{\partial(\ell^D)}.
\end{equation}
One then defines pressure as the left- or the right-hand side of
the above equation, times $(k_\mathrm{B}\,T)$. In the
noncommutative case, however, (\ref{04.141}) does not hold, and
one is faced with two different possible definitions for the
pressure. The first, based on the left hand side of
(\ref{04.141}), is
\begin{equation}\label{04.142}
P_1:=k_\mathrm{B}\,T\,\frac{\ln\mathcal{Z}}{V}.
\end{equation}
This is the pressure felt by something trying to move the
boundaries of the system. One has
\begin{equation}\label{04.143}
P_1=-\frac{k_\mathrm{B}\,T}{s}\,\int\d
U\;\ln\{1-s\,z\,\exp[-\beta\,E(U)]\}.
\end{equation}
To obtain the form of the second definition of pressure, based on
the right hand side of (\ref{04.141}), one notices that the
product $(V\,\d U)$ does not involve $\ell$. Hence,
\begin{equation}\label{04.144}
P_2=-\frac{k_\mathrm{B}\,T\,\ell^D}{s}\,\int\d
U\;\frac{\partial\ln\{1-s\,z\,\exp[-\beta\,E(U)]\}}{\partial(\ell^D)},
\end{equation}
which results in
\begin{equation}\label{04.145}
P_2=\int\d U\;n(U)\,\left[-\frac{\ell}{D} \,\frac{\partial
E(U)}{\partial\ell}\right].
\end{equation}
This is very similar to what obtained in the commutative limit,
when one notices that
\begin{equation}\label{04.146}
-\ell\,\frac{\partial}{\partial\ell}=p\,\frac{\partial}{\partial
p}\bigg|_{\hat k},
\end{equation}
where $\hat k$ in the superscript means differentiation with $\hat
k$ kept fixed.
\section{Asymptotic behaviors}
Let us consider a compact group. For such a group the energy
function is bounded. The minimum of energy is taken to be zero, by
convention. The maximum of energy ($E_\mathrm{max}$) is decreasing
in $\ell$, and tends to infinity as $\ell$ tends to zero. Another
length parameter is the so called thermal wavelength ($\lambda)$,
which is the proportional to the quantum (but commutative)
wavelength of a particle of energy $k_\mathrm{B}\,T$. Finally,
there is a length parameter associated to the density of
particles:
\begin{equation}\label{04.147}
\sigma:=\left(\frac{V}{\mathcal{N}}\right)^{1/D}.
\end{equation}
It is seen that of these three length parameters, $\ell$ is fixed,
$\lambda$ is a decreasing function of the temperature, and
$\sigma$ is a decreasing function of the density.
\subsection{High temperature limit}
In this case,
\begin{equation}\label{04.148}
\lambda\ll\sigma,\ell,
\end{equation}
from which
\begin{equation}\label{04.149}
\beta\,E_\mathrm{max}\ll 1.
\end{equation}
Putting this in the expressions for the number density, energy
density, and pressures, one arrives at
\begin{align}\label{04.150}
n(U)&=\frac{z}{1-s\,z},\\ \label{04.151}
\frac{\mathcal N}{V}&=\frac{z}{1-s\,z}\,\int\d U,\nonumber\\
&=\frac{z}{1-s\,z}\,\mathrm{vol}(G),\\
\label{04.152}
\frac{\mathcal E}{V}&=\frac{z}{1-s\,z}\,\int\d U\;E(U),\nonumber\\
&=\frac{\mathcal N}{V}\,\langle E\rangle,\\
\label{04.153}
P_1&=-\frac{k_\mathrm{B}\,T}{s}\,\ln(1-s\,z)\,\mathrm{vol}(G),\\
\label{04.154} P_2&=\frac{\mathcal
N}{V}\,\left\langle-\frac{\ell}{D}\, \frac{\partial
E}{\partial\ell}\right\rangle.
\end{align}
It is seen that at at this limit everything is temperature
independent, apart from $P_1$ which diverges linearly (in
temperature). It is also seen that the quantum behavior of the
system (which is manifested in the value of $s$) is
\textit{almost} not important (again apart from $P_1$). One can
define an effective fugacity $z_\mathrm{eff}$ as
\begin{equation}\label{04.155}
z_\mathrm{eff}:=\frac{z}{1-s\,z},
\end{equation}
to see that $s$ is actually eliminated from the expressions. By
\textit{almost} it is meant that for the case of fermions, the
density should not be greater than a critical limit:
\begin{equation}\label{04.156}
\sigma_\mathrm{cr}^{-D}:=\mathrm{vol}{G}.
\end{equation}
In fact this condition is not restricted to high temperatures.
\subsection{Low temperature limit}
In this case,
\begin{equation}\label{04.157}
\lambda\gg\sigma,\ell,
\end{equation}
from which
\begin{equation}\label{04.158}
\beta\,E_\mathrm{max}\gg 1.
\end{equation}
Here in all of the integrals involved in calculating the partition
function and thermodynamic quantities, only small values of $\hat
k$ have significant contributions. So one can use the asymptotic
forms of the integration measure based on (\ref{04.128}), and also
the commutative form of the energy function. This means that the
effects of noncommutativity are disappeared in this limit. The
quantum effects, however, are very strong. In fact in this limit
one encounters a highly degenerate but commutative gas.
\subsection{Moderate temperatures}
If $\sigma$ and $\ell$ are much different, there is a region for
temperature where $\lambda$ in between these two length scales and
much different from these. There arises two cases.
\subsubsection{Classical commutative behavior}
In this case,
\begin{equation}\label{04.159}
\ell\ll\lambda\ll\sigma.
\end{equation}
This case is, of course, possible only if
\begin{equation}\label{04.160}
\ell\ll\sigma.
\end{equation}
Here (\ref{04.158}) is satisfied, so that one can eliminate the
noncummutative parameter. For the effectively commutative system
resulted, as the thermal wavelength is much smaller than the
particle spacing, on has a nondegenerate (classical) gas.
\subsubsection{Quantum noncommutative behavior}
In this case,
\begin{equation}\label{04.161}
\sigma\ll\lambda\ll\ell.
\end{equation}
This case is, of course, possible only if
\begin{equation}\label{04.162}
\sigma\ll\ell.
\end{equation}
Here (\ref{04.149}) holds, and it is seen that both quantum and
noncommutative behaviors are pronounced. Note, however, that
(\ref{04.162}) and hence (\ref{04.161}) cannot be satisfied for
fermions, as for fermions (\ref{04.156}) shows that
\begin{equation}\label{04.163}
\sigma>2\,\pi\,\ell\,[\mathrm{vol_N}(G)]^{-1/D},
\end{equation}
where
\begin{equation}\label{04.164}
\mathrm{vol_N}(G):=(2\,\pi\,\ell)^D\,\mathrm{vol}(G),
\end{equation}
and the left hand side of (\ref{04.164}) (the dimensionless volume of the group) is
of the order of unit.
\subsection{Degenerate gases}
In this case,
\begin{equation}\label{04.165}
\sigma\ll\lambda.
\end{equation}
For bosons, the occurrence of Bose-Einstein condensation is
similar to the commutative case. It depends on whether the right
hand side of (\ref{04.137}) diverges for $z=1$ or not, and this is
determined by only the low momentum behavior of the energy
function and the integration measure. None of these depend on the
noncommutative parameter $\ell$. So exactly as it was in the
commutative case, the condition for the occurrence of
Bose-Einstein condensation is
\begin{equation}\label{04.166}
\lim_{p\to 0}\left(\frac{p}{E}\,\frac{\partial E}{\partial
p}\right)<D.
\end{equation}
The left hand is 1 for a relativistic gas and 2 for a
nonrelativistic gas. So one arrives at the familiar commutative
result that there is a Bose-Einstein condensation (for a
nonrelativistic gas) iff the dimension of the space is more than
2. The coexistence curve, however, does depend on the
noncommutative parameter $\ell$.

For fermions, it is seen from (\ref{04.137}) that the fugacity diverges when the density approaches
the critical density. One has
\begin{equation}\label{04.167}
\sigma^{-D}=\sigma_\mathrm{cr}^{-D}-\int\d U\;\frac{1}{1+z\,\exp[-\beta\,E(U)]},
\end{equation}
which results (up to leading order) in
\begin{equation}\label{04.168}
z=[\sigma_\mathrm{cr}^{-D}-\sigma^{-D}]\,\left\{\int\d U\;\exp[\beta\,E(U)]\right\}^{-1},
\end{equation}
showing that $z$ diverges like $(\sigma-\sigma_\mathrm{cr})^{-1}$.
It is also seen that as the density approaches the critical density, $P_1$ diverges:
\begin{equation}\label{04.169}
P_1=k_\mathrm{B}\,T\,[\mathrm{vol}(G)]\,\ln z
\end{equation}
(showing that $P_1$ diverges logarithmically), while $P_2$ tends to a finite value:
\begin{equation}\label{04.170}
P_2=\int\d U\;\left[-\frac{\ell}{D}\,\frac{\partial E(U)}{\partial\ell}\right].
\end{equation}

\subsection{Nondegenerate gases}
Here,
\begin{equation}\label{04.171}
\sigma\gg\lambda,
\end{equation}
so that the quantum behavior is not important. One then has
\begin{equation}\label{04.172}
\frac{\ln\Z}{V}=z\,\int\d U\;\exp[-\beta\,E(U)].
\end{equation}
The noncommutative behavior manifests itself at high temperatures,
where $\lambda$ becomes comparable to (or less than) $\ell$.
\section{The group SU(2), and nondegenrate gases}
As an example, let us obtain a closed form for the function (\ref{04.172}) for the
groups SU(2) and SO(3). To do so, one uses (\ref{04.133}) and (\ref{04.134}), and needs the
form of $E$. Examples are (\cite{fakE1,fakE2,skf,kfs,fsk})
\begin{equation}\label{04.173}
E=\begin{cases}
\displaystyle{\frac{4\,\hbar^2}{\ell^2\,m}\,\left(1-\cos\frac{\hat k}{2}\right),}&\mathrm{SU(2)}\\
\displaystyle{\frac{\hbar^2}{\ell^2\,m}\,(1-\cos\hat k),}&\mathrm{SO(3)}
\end{cases}.
\end{equation}
One then arrives at
\begin{equation}\label{04.174}
\ln\Z_\mathrm{SU(2)}=\frac{2\,V}{\ell\,\lambda^2}\,\exp\left(-\frac{2\,\lambda^2}{\pi\,\ell^2}\right)
\,\I_1\left(\frac{2\,\lambda^2}{\pi\,\ell^2}\right),
\end{equation}
and
\begin{equation}\label{04.175}
\ln\Z_\mathrm{SO(3)}=\frac{V}{\pi\,\ell^3}\,\exp\left(-\frac{\lambda^2}{2\,\pi\,\ell^2}\right)
\,\left[\I_0\left(\frac{\lambda^2}{2\,\pi\,\ell^2}\right)-
\I_1\left(\frac{\lambda^2}{2\,\pi\,\ell^2}\right)\right],
\end{equation}
where $\I_n$ is the modified Bessel function of order $n$.

\section{Conclusion}
Effects of noncommutativity on thermodynamic properties were explored
in spaces with commutation relations of a Lie algebra. In particular the
case of a Lie algebra corresponding to a compact Lie group was
investigated. In such cases the volume of the corresponding momentum space is finite.
A finite volume for the configuration space was introduced in terms of a Hermitian projection
the range of which covers the spaces of only certain representations in the regular representation.
The grandcanonical partition function of a system of identical free particles was then expressed,
in terms of the noncommutativity parameter, the fugacity, and the temperature. Regarding the concept of
pressure, it turned out that two ways are possible to define the pressure, as there are two ways
to change the volume of the system, either change the noncommutativity length parameter, or change
the largest representation entering the truncated (finite volume) system. While these give identical
results in the commutative limit, that is not the case for the noncommutative spaces. Different
asymptotic behaviors of physical quantities were explored. It was seen that there are three length
scales: the noncommutativity length scale, the thermal wavelength, and the mean particle separation
length. Of these, the last two are present in the commutative case. Quantum behavior is important when
the thermal wavelength is large, and noncommutativity is important when the noncommutativity length scale
is large. The effects of temperature and density on these were investigated. Finally, for the special
groups SU(2) and SO(3) the partition function was explicitly calculated in the nondegenrate limit
(where quantum effects are negligible).
\\[\baselineskip]
\textbf{Acknowledgement}:  This work was partially
supported by the research council of the Alzahra University.

\end{document}